# Collective scattering from quantum dots in a photonic crystal waveguide

J. Q. Grim,[1] I. Welland,[2] S. G. Carter,[1] A. S. Bracker,[1] A. Yeats,[1] C.S. Kim,[1] M. Kim,[3] K. Tran,[2] I. Vurgaftman,[1] T. L. Reinecke[1]

[1] *Naval Research Laboratory, 4555 Overlook Avenue SW, Washington, D.C. 20375, US*
[2] *NRC Research Associate at the Naval Research Laboratory, 4555 Overlook Avenue SW, Washington, D.C. 20375, USA*
[3] *Jacobs, Hanover, MD*

## Abstract

We demonstrate scattering of laser light from two InAs quantum dots coupled to a photonic crystal waveguide, which is achieved by strain-tuning the optical transitions of the dots into mutual resonance. By performing measurements of the intensity and photon statistics of transmitted laser light before and after tuning the dots into resonance, we show that the nonlinearity is enhanced by collective scattering. In addition to providing a means of manipulating few-photon optical nonlinearities, our approach establishes new opportunities for multi-emitter quantum optics in a solid-state platform.

## Introduction

An efficient photon-photon nonlinearity is a key resource needed to access and to use quantum information stored in light[1], [2], thus underpinning the operation of components such as single-photon switches and all-optical deterministic quantum logic gates [3]–[5]. An appealing approach to produce such a nonlinearity is with a coherent light-matter interface of quantum emitters such as atoms[6], [7] or solid-state 'artificial atoms'[8]–[12] coupled to a single-mode nanophotonic waveguide (WG). Operated as two-level systems, these emitters can be saturated by the single photon component of a resonant coherent input field, producing a giant optical nonlinearity at the quantum limit. This giant nonlinearity has recently been demonstrated by scattering weak coherent laser light from single semiconductor quantum dots (QDs) embedded in photonic crystal (PhC) WGs [8], [9], [11]. In this approach, a QD can manipulate the classical coherent input to generate a non-classical output, including exotic states of light such as energy-time entangled two-photon bound states[8], [9], [11], [13]. Waveguide quantum electrodynamics also offers a tantalizing platform for realizing multi-body quantum optics, with dispersion and modal engineering possibilities that are inaccessible to traditional cavity QED. This has generated significant recent theoretical and experimental interest from solid-state[10], [14]–[18], atomic[6], [14], [19], and superconducting qubit communities[20].

Collective scattering from multiple emitters (i.e. scattering single-frequency laser light from more than one emitter) could provide new opportunities for quantum information processing components such as single-photon switches and all-optical deterministic quantum logic gates [3]–[5], but it remains challenging for atomic systems and unexplored for solid-state emitter systems. For the former, despite progress in creating deterministic atom-WG interfaces[21], it remains difficult to localize and to efficiently couple atoms to WGs. It has been shown that this can be overcome by scattering from an ensemble of atoms weakly coupled to a waveguide, resulting in a collective enhancement of correlated photon pairs [6]. On the other hand, strong and deterministic WG coupling is routinely achieved with solid-state emitters such as QDs [8], [9], [11] and defects in diamond[10], [15], but the variation in optical transition frequencies of solid-state emitters has made the demonstration of multi-emitter quantum optics an ongoing challenge. Although there has been recent progress in demonstrating collective quantum phenomena such as superradiance through the emission of indistinguishable photons from multiple solid-state emitters coupled to the same WG mode[10], [14]–[18], collective scattering of laser light has not yet been demonstrated in the solid state.

In this Letter, we demonstrate collective scattering of laser light from two QDs coupled to a single mode PhC WG. This result is made possible by a strain-tuning technique that we recently developed [16], which allows the optical transitions of multiple QDs to be tuned into mutual resonance within the same PhC WG. For these experiments, a continuous-wave (cw) laser is transmitted through a photonic crystal WG with embedded InAs QDs. By tuning the laser on-resonance with a two-level QD optical transition, the single-photon component of the coherent laser field is reflected and correlated photon pairs (photon bound states) are transmitted [6], [8], [11]. The QD-WG coupling efficiency ($\beta$) is a key parameter that controls the size of the nonlinearity that can be observed in these experiments, with the magnitude of the transmission dip proportional to $1 - \beta$ for a transmitted weak input field resonant with a QD. We show that collective scattering produces a larger nonlinearity in the transmitted intensity as well as in the bunching in 2nd-order photon correlation measurements compared to individual QDs, with both effects arising from a higher probability of single photons scattering from two QDs than from one.

These concepts are illustrated schematically in Fig. 1(a) and with the calculations in Figs. 1(b),(c) using the input-output formalism developed in Refs. [11], [22] (see Supplemental Material [23] for details of the model). For a weak coherent input field and $\beta = 0.5$ for each QD, i.e. light transmitted through the waveguide has a 50% probability of scattering from each QD, as shown in Fig. 1(b). When the QDs are in resonance, the magnitude of the transmission dip $1 - T/T_0$ increases. The bunched photon statistics around $\tau = 0$ in the calculated $g^{(2)}(\tau)$ curves in Fig. 1(c) result from the preferential transmission of two-photon bound states compared to single photon states, due to the nonlinear interaction with the QDs at the few photon level[8], [11], [24]–[26]. Increasing the number of QDs to two essentially purifies the two-photon output, resulting in a larger bunching peak for two resonant QDs.

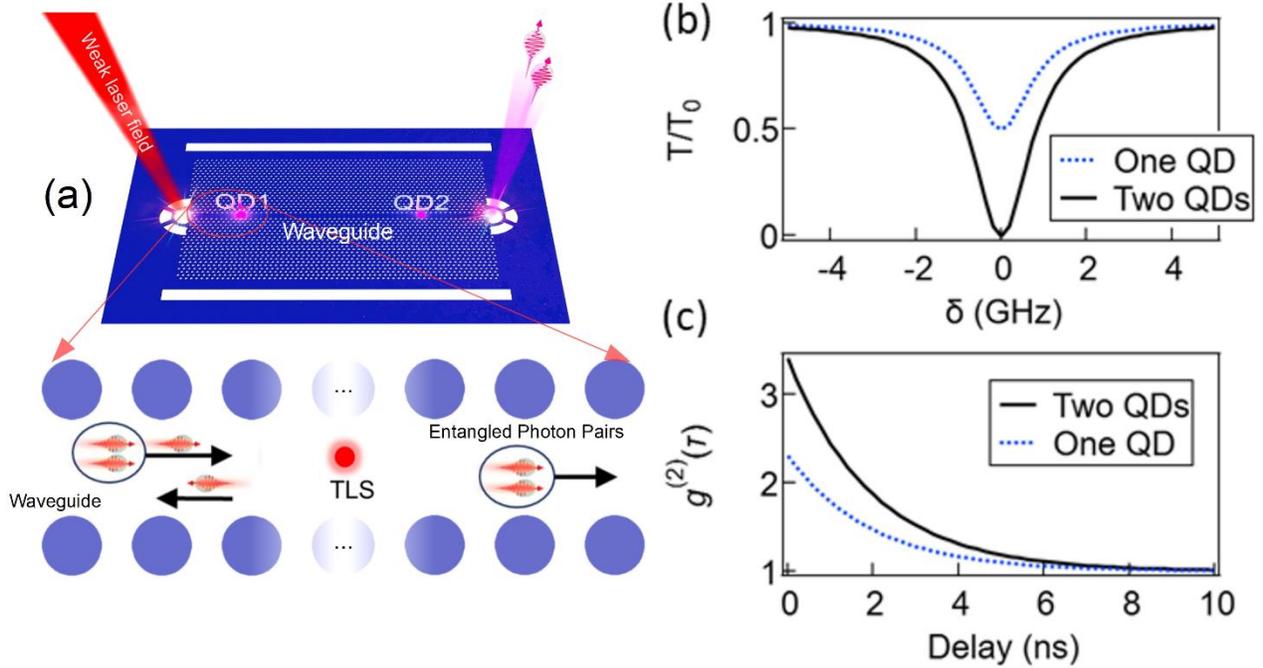

FIG. 1. (a) Photonic crystal waveguide containing embedded QDs and transmission experiment schematic. Calculated transmission (b) spectra and (c) 2nd-order autocorrelation $g^{(2)}(\tau)$ functions for one QD and two resonant QDs for a weak coherent input field and assuming an ideal waveguide (no end-facet reflections), with $\beta = 0.5$ and $\Gamma = 1$ GHz.

We use InGaAs QDs grown by molecular beam epitaxy embedded in GaAs PhC membrane waveguides with a vertical n-i-n-i-p heterostructure for deterministic control of the QD charge state[16]. Experiments were performed at 5.3 K with a 0.75 NA temperature-controlled objective in the vacuum space of a closed-cycle cryostat and a custom confocal µPL set up. We use the optically-active negative trion (X-) QD charge state, which has doubly-degenerate transitions from the electron spin ground states to X- states at zero magnetic field and can therefore be treated as a two-level system. The QDs are strain-tuned into resonance using a technique in which an above-gap laser is used to locally crystallize a thin HfO₂ film on the surface of the PhC membrane, allowing multiple QDs embedded in the same PhC WG to be brought into resonance[16]. The spatial positions of the QDs (labeled QD1 and QD2 throughout) were determined using an optical image of the sample and by scanning an above-gap laser over the waveguide while monitoring the PL. From this, we determine a QD1-QD2 distance of 4±0.3 µm.

We first characterize the photon statistics of the X- PL from two QDs excited by a 1353 meV non-resonant cw laser through the PhC WG mode, as shown in Fig. 2. We note that these $g^{(2)}(\tau)$ measurements of QD PL demonstrate the single- and indistinguishable-photon emission properties of the dots that were used for the collective scattering experiments shown in Figs. 3 and 4 and also serve as a comparison to previous work (Refs. [10], [15]–[17]). However, we emphasize that Fig. 2 and Fig. 4(b) display photon statistics that originate from different physical mechanisms. The former shows measurements of the photo-excited emission from the dots, and the latter shows measurements of transmitted laser light that resonantly scatters from the dots. In Fig. 2, we show the measured second-order autocorrelation function $g^{(2)}(\tau)$ using a Hanbury Brown-Twiss setup for three regimes in Fig. 2: i) single photon emission from one QD, ii) distinguishable emission from two- non-resonant QDs, iii) and indistinguishable emission from two resonant QDs. The single photon emission was measured by sending the emission from QD2 through a narrowband filter, resulting in antibunched $g^{(2)}(0) \approx 0.084$ statistics (black points in Fig. 2). After tuning QD1 to within 50 µeV of QD2, the emission from both QDs was sent through the same filter with $g^{(2)}(0) \approx 0.6$ (blue points in Fig. 2), as expected for two distinguishable photon sources[16]. The small discrepancy of the experimental data from the theoretical values of $g^{(2)}(0) = 0$ and 0.5, respectively, is predominantly due to background PL excited by the non-resonant laser. After tuning QD1 and QD2 into resonance, a bunching peak emerges with $g^{(2)}(0) = 0.86$ (red points in Fig. 2) due to the quantum interference of indistinguishable photons emitted by both QDs and is one signature of superradiance [15]–[17], [27]. The difference from the theoretical value of $g^{(2)}(0) = 1$ is due to a small spectral detuning between the QDs caused by the relatively high power of 2 µW used for the non-resonant 1353 meV excitation laser.

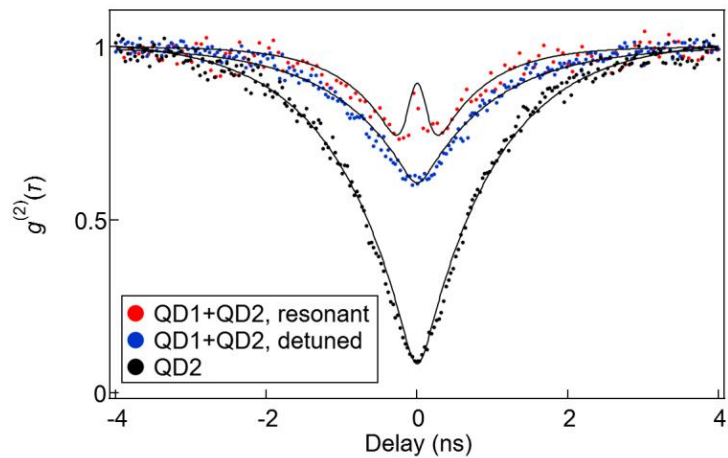

FIG. 2. 2nd-order autocorrelation measurements of the PL from i) one QD (black dots)), ii) two distinguishable QDs (blue dots), and iii) two indistinguishable QDs (red dots) excited by non-resonant 1353 meV laser light. $g^{(2)}(0) \to 0$ for one QD (QD2), $g^{(2)}(0) \to 0.5$ collecting emission from both QD1 and QD2 when they are detuned from one another, and $g^{(2)}(0) \to 1$ when QD1 and QD2 are resonant due to the quantum interference of indistinguishable photons in the waveguide[15]–[17], [27].

Next, we show the laser spectroscopy of these same two QDs, performed by transmitting a weak, tunable cw laser through the WG. In Fig. 3, the normalized transmission intensity $T/T_0$ is shown as a function of laser detuning (plotted relative to QD2) with QD1 and QD2 in the X$^-$ charge state. $T_0$ is the transmission intensity of the bare WG measured at an electrical bias far-detuned from the X$^-$ charge state, where the quantum dots do not absorb. The spectrum is shown before (red points) and after (black points) tuning QD1 into resonance with QD2, with transition linewidths of 0.48 GHz and 0.27 GHz for QD1 and QD2, respectively. The asymmetric line shapes originate from Fano interference caused by reflections from the PhC WG end facets that produce Fabry-Pérot modes[8], [11], and depend on the detuning from these modes (see Supplemental Material [23] for the PhC WG transmission spectrum).

We use the formalism of Ref. [28] to model the spectra in Fig. 3, and we account for spectral diffusion by running this calculation repeatedly with the spectral positions of each dot taken randomly and independently from a Gaussian distribution and averaging the results. For simplicity, we do not include the Fano effect in this model, but note that the change in the asymmetric lineshape before and after strain tuning is due a shift in the WG spectrum caused by the removal of condensed nitrogen from the PhC membrane due to laser heating. Using the radiative emission rate $\gamma_R/2\pi = 0.16$ GHz for both QD1 and QD2 determined from the measured $g^{(2)}(\tau)$ under non-resonant excitation (see Fig. 2), fits to the detuned QD1 and QD2 spectra in Fig. 3 yield the following parameters: $\beta_{QD1} = 0.14$, $\beta_{QD2} = 0.075$, and $\sigma_{SD,QD1} = 0.12$ GHz, $\sigma_{SD,QD2} = 0.065$ GHz, where $\sigma_{SD}$ is the broadening due to spectral diffusion. We use these parameters to calculate the transmission spectrum for zero detuning between the dots, and find excellent agreement with the two-dot experimental data (black points) for the magnitude of the transmission dip, as shown with the black dashed curve in Fig. 3. Further, the predicted increase in the width of the transmission dip for two dots in this model is due to an increased emission rate into the WG due to superradiance[28], which agrees very well with the measured linewidth of 0.64 GHz for the resonant dots. However, while the superradiant emission rate enhancement does not significantly change the magnitude of the predicted transmission dip, it does increase the bandwidth.

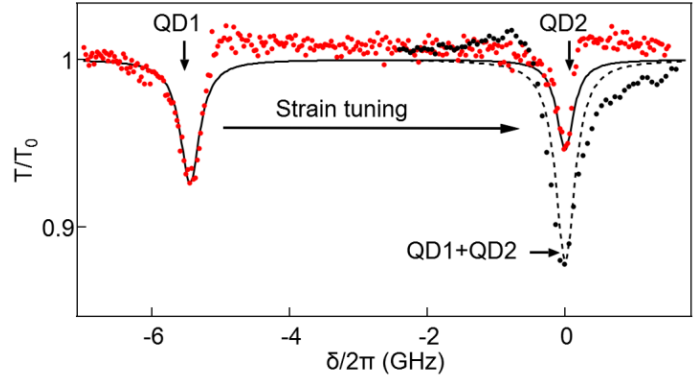

FIG. 3. The measured negative trion (X⁻) transmission spectrum for QD1 and QD2 as a function of detuning $\delta/2\pi$ from the excitation laser is shown prior to tuning (red points) as well as after strain-tuning QD1 into resonance with QD2 (black points). The parameters determined from fits to QD1 and QD2 (black line) were used to calculate the resonant QD spectrum (black dashed line).

We next demonstrate the nonlinearity of this system for individual and collective coherent scattering from QD1 and QD2. The following experiments were performed under the spectral conditions shown in Fig. 3. A cw laser was tuned on-resonance with each transition, and the transmitted power-dependent behavior is examined, providing a characteristic measure of saturation nonlinearity. Figure 4(a) plots the depth of the laser transmission dip for QD1 (blue squares), QD2 (red triangles), and QD1 and QD2 in resonance (black points) as a function of on-resonant laser power (measured before the objective). The average number of photons in the waveguide at the same time shown with the green axis, and is determined from the white light WG transmission spectrum (see Supplemental Material [23]). In each case, the magnitude of the transmission dip $(1 - T/T_0)$ decreases with increasing laser power due to saturation of the QDs, and eventually becomes transparent to the incoming field at high power (> 1 μW). This nonlinear change in transmission is larger for resonant QDs [QD1+QD2 in Fig. 4(a)] due to the increased depth of the transmission dip that originates from the increased probability of laser light scattering from two dots compared to one.

Insight into the quantum nature of this nonlinearity is gained from measuring the photon statistics of the transmitted laser. Figure 4(b) plots $g^{(2)}(\tau)$ for detuned individual QDs and for both QDs in resonance, all with $g^{(2)}(\tau = 0) > 1$ bunching, which originates from a greater probability of two photons being transmitted through the QDs compared to one photon[8], [9], [11]. The two-level QDs act to partially filter out single photon states. With the QDs in resonance, the bunching peak is larger, which results from a greater probability of filtering single photons due to collective scattering. We calculate $g^{(2)}(\tau)$ using the input-output formalism developed in Refs. [11], [22] [black lines in Fig. 4(b)] and the parameters obtained from fitting the transmission spectra shown in Fig. 3 (see Supplemental Material [23]). At high power, the QDs are effectively transparent due to saturation and the transmitted laser displays Poisson statistics, as shown in Fig. 4(b) for a laser power of 2 μW. In the Supplemental Material [23], we show additional examples of collective scattering from QDs with weaker coupling (lower $\beta$) as well as stronger (higher $\beta$) coupling efficiency to WGs. The same general features are observed, including an enhanced nonlinearity beyond the performance of the individual QDs, and a nonlinearity down to the single-photon level is observed for QDs with stronger coupling to the WG.

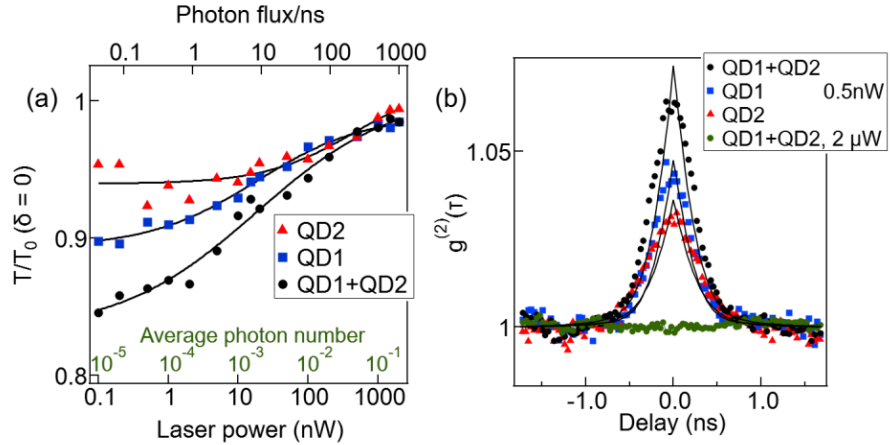

FIG. 4. Nonlinearity of resonant and non-resonant QDs in a waveguide as a function of transmitted laser power. (a) Saturation of transmission intensity on resonance for QD1, QD2, and QD1 + QD2 versus laser power (measured before the objective) under the spectral conditions shown in Fig. 3. The average photon number in the waveguide vs. power is shown with the green axis (after accounting for coupling losses). (b) 2nd-order autocorrelation function measurements of an on-resonant transmitted laser with a power of 0.5 nW before the objective. The bunching around zero delay ($g^{(2)}(0) > 1$) for QD1 (blue squares), QD2 (red triangles), and QD1 and QD2 in resonance (black circles) is due to the preferential transmission of two-photon bound states. The black lines are $g^{(2)}(\tau)$ calculations using parameters obtained from fitting the transmission spectra in Fig. 3. At high power (2 μW) above the saturation of the QDs, the Poissonian statistics of the input coherent laser field results in $g^{(2)}(0) = 1$, shown with green points.

Although the distance between dots was different in each of our examples and was not pre-selected, an important consideration is whether it has any effect on these experiments. For example, particular distances can result in cavity-like behavior with the QDs acting as quantum mirrors, which can produce a narrow transparency window in the transmission spectrum[20], [28]. However, in addition to requiring near-unity QD-WG coupling, we find that the spectral diffusion of the QDs used in this work makes the observation of this effect unlikely. Such observations may become possible with further advances in precise positioning of the QDs through site-controlled growth[29] within PhC membranes as well as the use of QDs with near-unity WG coupling[30] and transform-limited linewidths[31]. By extending this technique beyond two QDs and with somewhat better coupling to the WG, we expect that the transmission dip can go down much closer to zero, providing a highly nonlinear element at the single photon level.

This Letter demonstrates collective scattering from pairs of QDs coupled to PhC WGs, an achievement that opens opportunities for exploring the few-emitter regime that has been challenging to realize for both solid-state and atomic systems. This enables a systematic study of multi-emitter quantum optics and establishes the potential of this platform for realizing proposals of many-body states of light[32], [33]. The manipulation of quantum optical nonlinearities that has been made possible with this work has the potential to impact the creation of single-photon switches and all-optical deterministic quantum logic gates [3]–[5], the study of non-Markovian effects[34], and the formation of subradiant states for long-lived quantum memories[27], and open possibilities for the study and control of multi-photon dark states[20].

## Acknowledgements

This work was supported by the US Office of Naval Research. I. W. and K. T. are NRC Research Associates at the US Naval Research Laboratory.

# Supplemental Material for

## Collective scattering from quantum dots in a photonic crystal waveguide

J. Q. Grim,[1] I. Welland,[2] S. G. Carter,[1] A.S. Bracker,[1] A. Yeats,[1] C.S. Kim,[1] M. Kim,[3] K. Tran,[2] I. Vurgaftman,[1] T. L. Reinecke[1]

[1] Naval Research Laboratory, 4555 Overlook Avenue SW, Washington, D.C. 20375, US
[2] NRC Research Associate at the Naval Research Laboratory, 4555 Overlook Avenue SW, Washington, D.C. 20375, USA
[3] Jacobs, Hanover, MD

## I.  Model of resonant scattering from two emitters

The nonlinear behavior observed when multiple QDs are tuned on resonance is ultimately controlled by the differing interactions between the single and multi-photon components of the incident photon field and the QD. The correlation functions $g^{(1)}$ and $g^{(2)}$ are related to the S matrix elements for the single and two photon component respectively [1]. Therefore, if the $g^{(2)}$ correlation function computed via a master equation for two dots increases for $g^{(2)}(0)$, one may infer that the suppression of the one-photon component accounts for this increase. The Hamiltonian for our system is given by [1],

$$H = -(\omega - \omega_0)\Sigma_i \sigma_i^+ \sigma_i^- + i\Omega(t)\Sigma_i(\sigma_i^+ - \sigma_i^-) \tag{1}$$

where $\sigma_i^{\pm}$ is the raising/lowering operator for the respective quantum dots, $\omega_0$ is the resonant frequency of the dot (assumed identical in both dots when tuned on resonance), and $\Omega$ is the driving strength. This model assumes that the dots are completely uncorrelated with one another as they are uncoupled, and moreover retarded effects are neglected. The master equation associated with this Hamiltonian is, $\dot{\rho} = i[H,\rho] + \gamma_{tot}D[\sigma^-]\rho + 2\gamma_d D[\sigma^+\sigma^-]\rho$, where $D[x]\rho \equiv x\rho x^\dagger - (x^\dagger x \rho - \rho x^\dagger x)/2$ is the Lindblad superoperator, $\gamma_d$ is the pure dephasing rate, and $\gamma_{tot}$ is the total radiative decay. This equation may be solved utilizing the package QuTip [2], [3]. We compute the necessary correlation functions by invoking the input-output relations derived in [4],

$$a_{out} = za_{in} - i\frac{z}{|z|^2}\sqrt{\frac{\beta_{tot}}{2}}\, \Sigma_i \sigma_i^-(t)e^{\frac{i\omega(x-x_i)}{c}} \tag{2}$$

where $z \equiv \frac{1}{1+i\zeta}$ captures the reflection coefficient of the waveguide in the absence of the emitter, and therefore the Fano effect. A Markovian factor $e^{\frac{i\omega(x-x_i)}{c}}$ captures the spatial dependence of the coupling of the dot to the incident photons. We neglect this factor for the rest of our analysis on account of spectral wandering, as any effect it would have is averaged over due to the wide bandwidth.

Fits to the transmission data in Fig. 3(a) were performed utilizing a Metropolis-Hastings Markov Chain Monte Carlo approach. The intensity is given by,

$$I = \frac{\langle a_{out}^\dagger(t) | a_{out}(t) \rangle_{ss}}{|z|^2 |\alpha|^2 F},$$ (3)

where $F$ is the photon flux. Similarly, we may compute the second order correlation function via

$$G^{(2)} = \frac{\langle a_{out}^\dagger(t) a_{out}^\dagger(t+\tau) a_{out}(t+\tau) a_{out}(t) \rangle}{|z|^4 |\alpha|^4 F^2}.$$ (4)

We fit separate parameters for each dot individually, then utilize both sets of parameters to compute the intensity of the two dot system as well as the second-order correlation function, using the expressions for the correlation functions derived in Ref. [1]. In the two dot case, we use the generalized input-output relations of Ref. [5]. Consistent with experiment, we observe a transmission intensity decrease on resonance for two dots, and an increase of $g^{(2)}(0)$. The Fano model used when two dots are in resonance in the waveguide does not account for reflections between dots and essentially treats the dots as a single entity. Finally, the narrower width of the predicted two dot transmission spectrum shown in Fig. S1 compared to the model used in Fig. 3 in the main text is because superradiance is not included. However, both models predict similar magnitudes of the transmission dips.

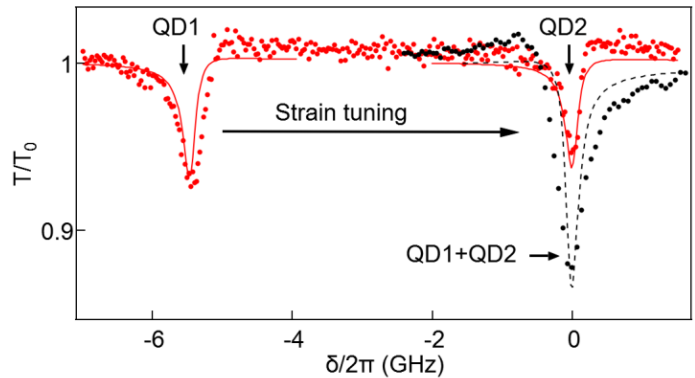

FIG.S1. Measured negative trion (X⁻) transmission spectrum for QD1 and QD2 as a function of detuning $\delta/2\pi$ from the excitation laser is shown prior to tuning (red points) as well as after strain-tuning QD1 into resonance with QD2 (black points) as shown in Fig. 3 of the main text. The red lines show fits to QD1 and QD2 using the input-output formalism presented above, with $\beta_{QD1}$ = 0.09, $\beta_{QD2}$ = 0.06, $\sigma_{SD,QD1}$ = 0.27 GHz, $\sigma_{SD,QD2}$ = 7.8e-4 GHz, $z_{QD1}$ = -0.15, and $z_{QD2}$ = -2.08. Pure dephasing was assumed to be negligible. These parameters were used to calculate the resonant QD spectrum (black dashed line) as well as the $g^{(2)}(\tau)$ curves in Fig. 4(b) in the main text.

## II. Photonic crystal waveguide transmission spectrum

The laser transmission spectrum of QD1 and QD2 in resonance is plotted relative to the white light transmission spectrum of the WG in Fig. S2. The fringes in the WG spectrum are due to reflections from the end facets, which results in the Fano lineshapes of the QD transmission spectra. From the WG spectrum, we estimate a group index of $n_g \approx 7.5$, resulting in an average of 1.9 photons at the same time for 1 μW of laser power in the waveguide. The group index is calculated using $n_g = \lambda^2 \, / \, 2L\Delta\lambda$, where $\lambda$ is the QD emission wavelength, $L$ is the length of the waveguide, and $\Delta\lambda$ is the oscillation period resulting from Fabry-Perot interference. We estimate the average photon number from $\langle n \rangle = \eta_c F \tau_{ph} = \frac{\eta_c F L n_g}{c \, (1-R)}$, where $\eta_c$ is the coupling efficiency, $L$ is the length of the waveguide, and $R = 17\%$ is the estimated reflectivity of the waveguide end facets from the fringes in the white light transmission spectrum

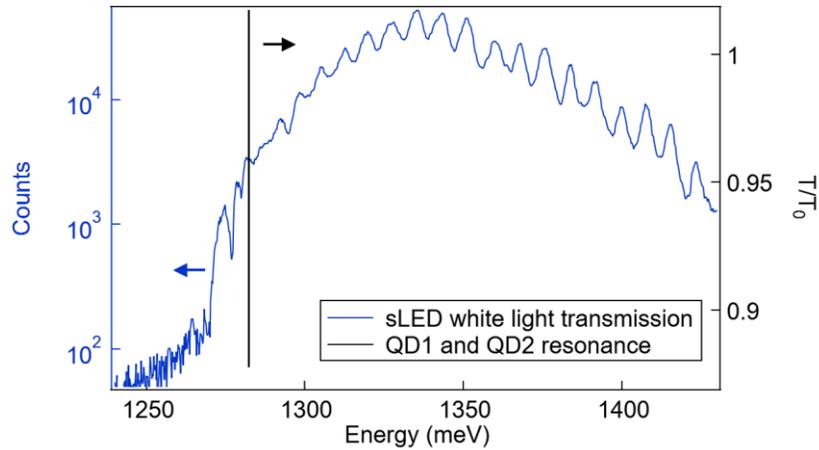

FIG.S2. White light waveguide transmission spectrum and laser transmission spectrum of QD1 and QD2 in resonance.

## III. Additional examples of QD collective scattering

Here, we present additional examples of collective scattering from QDs in different WGs. In the first example (Figs. S3 and S4), the QDs are weakly coupled to the waveguide with $\beta < 0.05$ for both QDs. Fig. S3(a) shows $g^{(2)}(\tau)$ measurements under non-resonant excitation similar to the data shown for different QDs in the main text under the following conditions: i) one-QD single photon emission, ii) two-QD non-resonant emission, iii) and two-QD resonant emission. The single photon emission from one QD was measured by sending the emission from QD1a through a narrowband filter, resulting in antibunched $g^{(2)}(0) \rightarrow 0$ statistics. After tuning QD1a to within 50 μeV of QD2a, the emission from both QDs was sent through the same filter with $g^{(2)}(0) \rightarrow 0.5$, as expected for two distinguishable photon sources.[6] After tuning QD1a and QD2a into resonance, $g^{(2)}(0) \rightarrow 1$ due to the quantum interference of indistinguishable photons emitted by both QDs and is a signature of superradiance [6]–[9]. The laser transmission spectra of the X$^-$ charge states is shown in Fig. S3(b) prior to tuning the QDs into resonance, and the spectrum after tuning the QDs into resonance is plotted relative to the WG white light transmission spectrum in Fig. S3(c).

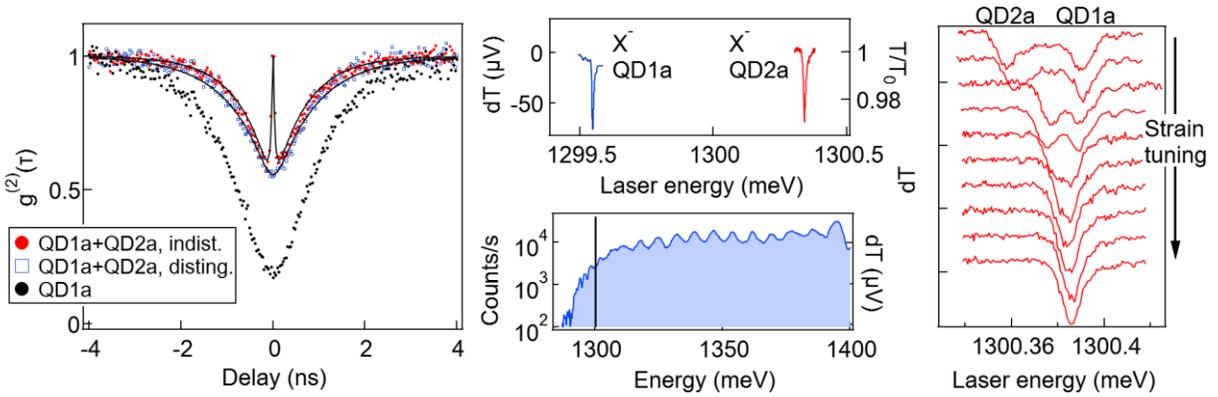

FIG. S3. (a) 2nd-order autocorrelation measurements for non-resonant (1345 meV) excitation through the waveguide for one QD, two distinguishable QDs, and two indistinguishable QDs. $g^{(2)}(0) \rightarrow 0$ for one QD, $g^{(2)}(0) \rightarrow 0.5$ collecting emission from both QD1a and QD2a when they are detuned from one another, and $g^{(2)}(0) \rightarrow 1$ when QD1a and QD2a are resonant due to the quantum interference of indistinguishable photons in the waveguide[6]–[9]. The inset schematic shows the approximate positions of the QD1 and QD2 within the PhC WG. (b) Laser transmission spectra of the negative trion (X$^-$) QD1a and QD2a prior to tuning, (c) white light transmission spectrum of the waveguide compared to the resonance of the two QDs, (d) laser transmission spectra taken after strain tuning steps showing QD2 tuned into resonance with QD1.

Despite the weak QD-WG coupling ($T/T_0 < 3\%$ for QD1a and QD2a), we observe an enhanced nonlinear response from collective coherent scattering compared to the response of the individual QDs, as shown in Fig. S4(a). The critical photon flux per ns characterizing the saturation nonlinearity is ~275, ~217, and ~61 for QD1a, QD2a, and QD1a + QD2a, respectively, demonstrating the impact of collective scattering. Similar to the example shown in the main text, the bunching around $g^{(2)}(\tau = 0)$ also increases for these QDs.

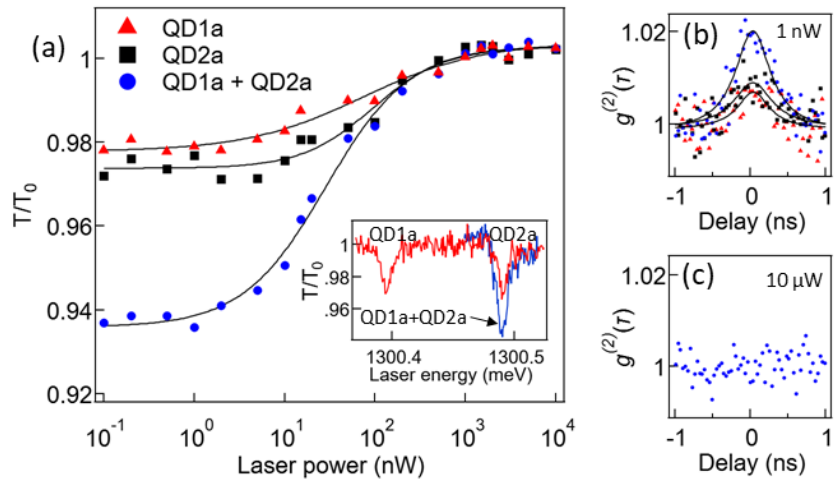

FIG. S4. Nonlinearity of resonant and non-resonant QDs in a waveguide. (a) Nonlinear saturation of transmission on resonance for QD1a, QD2a, and QD1a + QD2a versus laser power under the conditions shown in the inset. Inset: Laser transmission spectra of QD1a and QD2a detuned and after tuning QD1a into resonance with QD2a. The spectra are normalized with the bare-waveguide transmission spectrum $T_0$ (i.e. transmission at an electrical bias far-detuned from the QD1a and QD2a X⁻ charge stability region. The quantum nature of the nonlinearity is revealed by (d) high- and (e) low-power 2nd-order autocorrelation function measurements. The bunching around zero delay ($g^{(2)}(0) > 1$) in (e) is due to the preferential transmission of two-photon bound states. At high power above the saturation of the QDs, the Poissonian statistics of the input coherent laser field results in $g^{(2)}(0) = 1$.

In the next example, we show collective scattering from QDs with much stronger coupling to a PhC WG. The stronger coupling is partly due to the spectral proximity of the QDs to the PhC WG photonic band edge. Fig. S5(a) displays the X⁻ spectrum of both (resonant) QDs relative to the WG transmission spectrum, with the QD spectra prior to tuning shown in the inset. The $g^{(2)}(\tau)$ measurements of the transmitted laser shown in Fig. S4(c) were performed on-resonance for the spectral conditions shown in Fig. S5(b) for QD1b and QD2b prior to tuning. The increased probability of transmitting two-photon bound states results in a bunching peak with $g^{(2)}(0) \sim 1.95$.

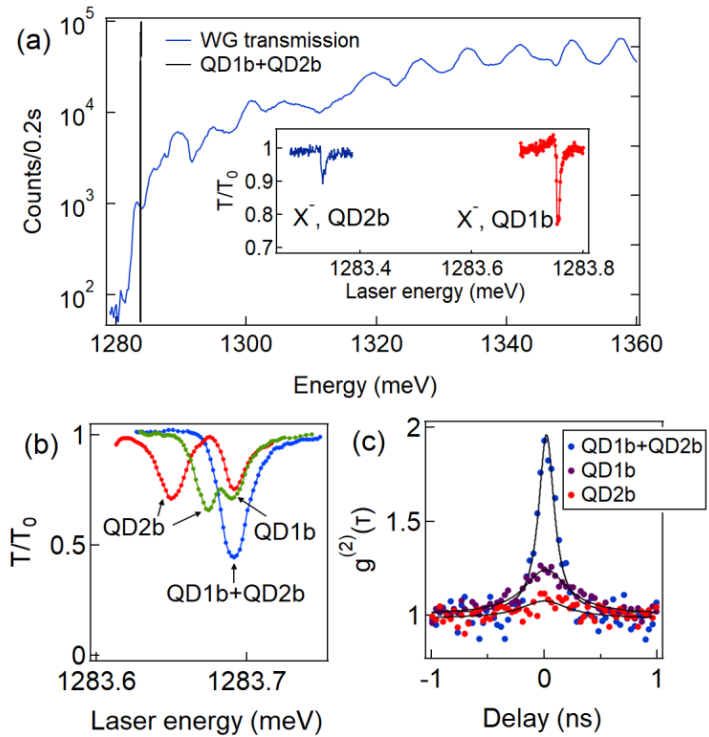

FIG. S5. Waveguide 2. (a) Waveguide transmission spectrum (blue) compared to two QDs tuned into resonance. Inset: transmission spectra of two QDs prior to tuning into resonance. (b) Transmission spectra of QD1b and QD2b after the final few tuning steps. (c) Measured $g^{(2)}(\tau)$ for QD1b and QD2b at resonance before and after tuning into resonance.

The larger than expected increase in the bunching peak shown in Fig. S5(c) is due to a shift in the WG transmission spectrum before and after tuning. The $g^{(2)}(\tau)$ measurements of the individual QDs were performed under the spectral conditions shown in the inset of Fig. S5(a) prior to tuning. Condensed nitrogen that was initially present on the surface of the PhC membrane was gradually removed by the laser heating used to tune the QDs, which caused a progressive blue-shift of the WG transmission spectrum. As a consequence, the QDs were closer to the WG band edge after they were tuned into resonance which resulted in an increase in the transmission amplitude. The blue-shift of the WG band edge saturated as condensed nitrogen was removed from the surface from repeated laser-heating exposures, and the transmission spectra of QD1b and QD2b after the final few tuning steps in Fig. S5(b) were measured under the same WG conditions.

### IV.        Sample structure and experimental details

InGaAs QDs were grown by molecular beam epitaxy within a membrane diode heterostructure on an n-doped GaAs substrate. The diode allows injection of electrons into the QD. The heterostructure consisted of a 0.5 μm Si-doped (n-type) GaAs buffer, 950 nm of Si-doped $Al_{0.7}Ga_{0.3}As$, 30 nm Si-doped n-type GaAs; 30 nm undoped GaAs; InGaAs QDs grown with a partial cap (nominal 2.7 nm) followed by a brief anneal; 71 nm undoped GaAs; 10 nm n-type Si-doped GaAs; 10 nm undoped GaAs; 30 nm p-type Be-doped GaAs. The intermediate n-type layer in this n-i-n-i-p diode reduces the forward bias required to charge the QD, avoiding high currents through the device. The waveguide patterns were produced using electron-beam lithography and $SiCl_4$-based inductively coupled plasma. The $Al_{0.7}Ga_{0.3}As$ was undercut with 10% hydrofluoric acid. Electrical contacts were made with indium to the p-type layer and to the back of the n-doped substrate. The 15-μm-long GaAs photonic crystal membrane WGs were 184 nm thick, and were terminated with semicircular grating couplers. The photonic crystal membranes consisted of a triangular lattice of holes with a lattice constant of 244 nm.

The positions of the QDs are random throughout the sample, and we identify QDs that couple to the WG by sending an above-gap laser through one grating coupler and collecting the photoluminescence (PL) from the opposite coupler to a spectrometer/CCD in the configuration shown in Fig. 1(a). The location of specific QDs is determined using an optical image of the sample and by scanning an above-gap laser along the waveguide while monitoring the PL.

## Acknowledgements

This work was supported by the US Office of Naval Research. I. W. and K. T. acknowledge the support of the NRC Research Associateship Program at the US Naval Research Laboratory.